\documentstyle[emulateapj]{article}
\begin{document}

\title{NEW EVIDENCE FOR THE UNIFIED SCHEME OF BL LAC OBJECTS AND FRI RADIO GALAXIES}

\author{J.M. Bai\altaffilmark{1,2} and Myung Gyoon Lee\altaffilmark{1}}

\altaffiltext{1}{Astronomy Program, SEES, Seoul National University, Seoul 151-742, Korea;\\
(jmmbai@astro.snu.ac.kr; mglee@astrog.snu.ac.kr)}
\altaffiltext{2}{Yunnan Astronomical Observatory, The Chinese Academy of Science,
Kunming 650011, China}

\begin{abstract}

In this paper, we collect radio and X-ray observations for most
Fanaroff-Riley I (FRI) radio galaxies in the Zirbel-Baum radio galaxy sample,
and investigate the distribution of the radio-to-X-ray
effective spectral index, $\alpha_{rx}$,
to test the unified scheme of BL Lac objects and FRI radio galaxies. 
It is found that the range of $\alpha_{rx}$ for FRI radio galaxies is almost the same as that for 
BL Lac objects, that  the distribution of $\alpha_{rx}$ probably peaks at the same position
as BL Lac objects, and  that the distribution of $\alpha_{rx}$ for FRIs
is similar to that for BL Lac objects. These suggest that 
there exist two subclasses of FRI radio galaxies: one is
HBL-like, and the other is LBL-like, corresponding to high-energy-peaked (HBL) and low--energy-peaked
(LBL) BL Lac objects, respectively. This result is consistent with previous
VLA observations, and supports 
the unified scheme of BL Lac objects and FRI radio galaxies.
\end{abstract}


\section{INTRODUCTION}

Among active galactic nuclei (AGNs), BL Lac objects are the most extreme
class characterized by strong and rapid variability, high polarization,
and weak emission lines. These extreme properties are generally interpreted
as a consequence of nonthermal emission from a relativistic jet oriented
close to the line of sight (Blandford \& K$\ddot{o}$nigl 1979).
Implicit in this hypothesis is that there exist many objects intrinsically
identical to BL Lac objects, with the relativistic jets oriented at large angle to the
line of sight (in the plane of the sky or perpendicular to our line of sight).
While BL Lac objects are dominated by relativistically beamed emission
from the jets, these objects are most likely dominated by unbeamed,
isotropic emission and appear to be very different. These objects constitute
the so-called parent population of BL Lac objects, and the unification 
in this way is usually called the ``unified scheme" (Antonucci \& Ulvestad 1985).
It is generally believed that the parent population of BL Lac objects is 
Fanaroff-Riley class I (FRI, Fanaroff \& Riley 1974) radio galaxies
(Browne 1983; Wardle, Moore \& Angel 1984).
There is growing evidence for bulk relativistic motion in the jets of blazars 
(BL Lac objects and OVV quasars), and the relativistic beaming model has been
widely accepted (Ghisellini et al. 1993). However, almost all arguments for
beaming can probably be explained in other ways as well (Urry \& Padovani
1995). The strongest challenge for this model remains in testing the unified
scheme (Urry \& Padovani 1990). Numerous studies have found general agreement
with the unified scheme of BL Lac objects and FRI radio galaxies, including
studies of 1) unbeamed properties of
BL Lac objects and FRI radio galaxies, such as extended radio emission,
narrow emission lines, host galaxies and environments, 2) luminosity
functions of the parent and beamed populations in different bands,
3) correlation between core-dominance parameter
(the ratio between core and extended radio flux) and beamed properties
(for a review see Antonucci 1993; Urry \& Padovani 1995).
Recently, Hubble Space Telescope ($HST$)
observations of host galaxies of BL Lac objects (Falomo et al. 1997,
Urry et al. 1999) and of the optical cores of FRI radio galaxies
(Chiaberge et al. 1999, 2000; Capetti \& Celotti 1999) also supported this unified scheme.
Furthermore, superluminal motion confirmed in M87 by $HST$ optical
observation (Bireta et al. 1999) and in B2 1144+35 by VLBI observation
(Giovannini et al. 1999) provided strong evidence to this unified scheme.

Early multiwavelength observations showed that emission from BL Lac objects
is dominated by synchrotron component with the peak lying in the IR to X-ray
energy range. In the 1990's, with the observations by Compton Gamma-Ray
Observatory and ground-based Cherenkov telescopes it was found that BL Lac
objects emit enormous power in rapidly variable high energy gamma rays (GeV
and TeV), indicating a second peak in their overall spectral energy
distribution (SED; Ulrich, Maraschi and Urry 1997; Fossati et al. 1998). 
A practical way to parameterize the different SEDs is to use the
radio-to-X-ray effective spectral index $\alpha_{rx}$. 
According to whether $\alpha_{rx}$ is greater than or less than 0.75
(Padovani \& Giommi 1995) BL Lac objects are divided into two subclasses,
the low-energy-peaked BL Lac objects (LBLs) which have synchrotron peaks in
IR/optical and high-energy-peaked BL Lac objects (HBLs) which have
synchrotron peaks in UV/soft X-ray. Most radio-selected BL Lac objects (RBLs)
are LBLs, while most X-ray-selected BL Lac objects (XBLs) are HBLs.
Further studies found that
the SED difference between LBL and HBL cannot be explained in terms of
different viewing angles alone, suggesting that there must be some
intrinsic differences that cause different SEDs between LBL and HBL
(Sambruna et al. 1996, 1999; Georgranopoulos \& Marscher et al. 1998;
Fossati et al. 1998). 

If FRI radio galaxies are, as believed today, misaligned BL Lac objects,
it is expected that the jet-related nonthermal emission from FRI radio
galaxies also has a double-peak structure in the SED, and that there exist two
intrinsically different subclasses of FRI radio galaxies: one is LBL-like
with $\alpha_{rx}>0.75$, and the other is HBL-like with $\alpha_{rx}<0.75$,
corresponding to the parent populations for LBLs and HBLs, respectively.
Therefore, the double-peak characteristics of the SED of the jet-related
emission and the existence or nonexistence of two
subclasses of FRI radio galaxies (or more strictly speaking, the distribution
of $\alpha_{rx}$ in FRI radio galaxies) can be used to test the unified
scheme of BL Lac objects and FRI radio galaxies.

There are some difficulties at present to test the unified scheme in terms
of the double-peak characteristics of the SED, because the detectors available
are not yet sensitive enough to detect high-energy emission in most of FRI
radio galaxies. In this paper we
collect X-ray and radio observations available in the literature
for most of FRIs in the Zirbel-Baum sample (Zirbel \& Baum 1995), and
investigate the distribution of $\alpha_{rx}$ for the sample
to test the unified scheme of BL Lac objects and FRI radio galaxies. 

\begin{deluxetable}{lllrrrrr}
\tablewidth{0pc}
\tablecaption{The sample of FRI radio galaxies}
\tablehead{
\colhead{IAU Name}    & \colhead{Other Name} & \colhead{Redshift} &
\colhead{$F_{5GHz}$} & \colhead{Ref.} &
\colhead{$F_{1keV}$} & \colhead{Ref.} &
\colhead{$\alpha_{rx}$}\\
\colhead{(1)} & \colhead{(2)} & \colhead{(3)} & \colhead{(4)(mJy)} &
\colhead{(5)} & \colhead{(6)(nJy)} & \colhead{(7)} & \colhead{(8)}}

\startdata
  0055$-$01 & 3C29   & 0.0448&     93.0&  1  &        11.1  &  2  & 0.82 \nl
  0104$+$32 & 3C31   & 0.0167&    140.0&  3  &        63.7  &  4  & 0.74 \nl
  0123$-$01 & 3C40   & 0.0180&    100.0&  1  &        58.5  &  2  & 0.73 \nl
  0219$+$42 & 3C66B  & 0.0215&    182.0&  5  &       141.0  &  4  & 0.71 \nl
  0255$+$05 & 3C75   & 0.0232&     39.0&  1  &       345.1  &  2  & 0.58 \nl
  0300$+$16 & 3C76.1 & 0.0320&     10.0&  6  &        33.0* &  7  & 0.61 \nl
  0305$+$03 & 3C78   & 0.0288&    964.0&  1  &       698.6  &  2  & 0.72 \nl
  0314$+$41 & 3C83.1B& 0.0255&     40.0&  8  &        27.0  &  4  & 0.72 \nl
  0316$+$41 & 3C84   & 0.0172&  59600.0&  9  &      1300.0  &  4  & 0.92 \nl
  0320$-$37 & ForA   & 0.0050&     26.0&  1  &       210.1  &  2  & 0.58 \nl
  0427$-$53 & PKS    & 0.0380&     57.0&  1  &       116.9  &  2  & 0.66 \nl
  0430$+$05 & 3C120  & 0.0330&   3458.0&  1  &      9065.5  &  2  & 0.65 \nl
  0453$-$20 & B2     & 0.0350&     40.0&  1  &        48.7  &  2  & 0.69 \nl
  0620$-$52 & PKS    & 0.0510&    260.0&  1  &       488.5  &  2  & 0.66 \nl
  0625$-$35 & OH-342 & 0.0550&    600.0&  1  &      1327.7  &  2  & 0.65 \nl
  0722$+$30 & B2     & 0.0191&     51.0&  10  &   $<$    20.0  &  11  & 0.75 \nl
  0800$+$24 & B2     & 0.0433&      3.0&  10  &   $<$     3.0  &  11  & 0.70 \nl
  0915$-$11 & PKS    & 0.0540&    217.0&  1  &      7314.8  &  2  & 0.50 \nl
  1040$+$31 & B2     & 0.0360&     55.0&  10  &        16.0  &  11  & 0.77 \nl
  1142$+$19 & 3C264  & 0.0208&    200.0&  12  &       486.0  &  4  & 0.65 \nl
  1144$+$35 & B2     & 0.0630&    250.0&  3  &       170.0  &  11  & 0.72 \nl
  1216$+$06 & 3C270  & 0.0060&    308.0&  1  &       147.5  &  2  & 0.74 \nl
  1222$+$13 & 3C272.1& 0.0031&    180.0&  10  &        26.0  &  4  & 0.81 \nl
  1228$+$12 & 3C274  & 0.0043&   4000.0&  10  &      1080.0  &  4  & 0.78 \nl
  1251$-$12 & 3C278  & 0.0150&     88.0&  1  &        79.3  &  2  & 0.71 \nl
  1254$+$27 & B2     & 0.0246&      2.3&  11  &        25.0  &  11  & 0.57 \nl
  1256$+$28 & B2     & 0.0224&      2.0&  3  &   $<$     2.0  &  11  & 0.70 \nl
  1317$+$33 & B2     & 0.0379&      8.0&  11  &         7.0  &  11  & 0.71 \nl
  1318$-$43 & NGC5090& 0.0110&    580.0&  1  &       250.5  &  2  & 0.75 \nl
  1322$+$36 & B2     & 0.0175&    150.0&  3  &   $<$    10.0  &  11  & 0.85 \nl
  1322$-$42 & CEN A  & 0.0020&   6984.0&  1  &       488.5  &  2  & 0.85 \nl
  1333$-$33 & PKS    & 0.0130&    297.0&  1  &       179.5  &  2  & 0.73 \nl
  1346$+$26 & B2     & 0.0633&     53.0&  10  &        36.0  &  11  & 0.72 \nl
  1414$+$11 & 3C296  & 0.0237&     77.0&  10  &        57.8  &  4  & 0.72 \nl
  1422$+$26 & B2     & 0.0370&     25.0&  10  &         4.0  &  11  & 0.80 \nl
  1511$+$26 & 3C315  & 0.1080&    150.0&  13  &        42.0* &  7  & 0.74 \nl
  1514$+$07 & 3C317  & 0.0342&    391.0&  1  &      5167.4  &  2  & 0.55 \nl
  1553$+$24 & B2     & 0.0426&     53.6&  14  &        11.0  &  11  & 0.79 \nl
  1610$+$29 & B2     & 0.0313&      6.0&  10  &        12.0  &  11  & 0.66 \nl
  1621$+$38 & B2     & 0.0310&     50.0&  10  &        16.0  &  11  & 0.76 \nl
  1626$+$39 & 3C338  & 0.0298&    105.0&  3  &        17.5  &  4  & 0.80 \nl
  1648$+$05 & 3C348  & 0.1540&     10.0&  1  &       790.5  &  2  & 0.45 \nl
  1836$+$17 & 3C386  & 0.0180&     14.0&  15  &        30.0* &  7  & 0.63 \nl
  1839$-$48 & PKS    & 0.1120&    163.0&  1  &       229.6  &  2  & 0.68 \nl
  1855$+$37 & B2     & 0.0552&    100.0&  10  &        35.0  &  11  & 0.76 \nl
  1954$-$55 & PKS    & 0.0600&     50.0&  1  &       204.6  &  2  & 0.62 \nl
  2058$-$28 & PKS    & 0.0380&     63.0&  1  &       116.9  &  2  & 0.66 \nl
  2152$-$69 & PKS    & 0.0270&    400.0&  1  &       903.2  &  2  & 0.65 \nl
  2212$+$13 & 3C442A & 0.0263&      2.0&  10  &   $<$    15.0  &  4  & 0.59 \nl
  2229$+$39 & 3C449  & 0.0171&     37.0&  10  &        16.0  &  4  & 0.75 \nl
  2335$+$26 & 3C465  & 0.0293&    270.0&  3  &        65.8  &  4  & 0.78 \nl
\tablecomments{The X-ray flux densities with $^{\ast}$ are at 2keV.}
\tablerefs{(1) Morganti et al. 1993;  (2) Siebert et al. 1996;
(3) Giovannini et al. 1990; (4) Hardcastle \& Worrall 1999;
(5) Leahy et al. 1986; (6) Macklin 1983; (7) Fabbiano et al. 1984;
(8) O'Dea \& Owen 1985; (9) Noordam \& de Bruyn 1982;
(10) Giovannini et al. 1988; (11) Canosa et al. 1999;
(12) Gavazzi et al. 1981; (13) H$\ddot{o}$gbom 1979;
(14) Fanti et al. 1987; (15) Strom et al. 1978.}
\enddata
\end{deluxetable}

\section{THE SAMPLE}

Our sample, as listed in Table 1, comprises most of FRI radio galaxies in the
sample of radio-loud elliptical galaxies compiled by Zirbel \& Baum (1995).
The Zirbel-Baum sample is the largest known
sample of radio-loud elliptical galaxies (excluding quasars).
The radio
luminosity ranges from the ``radio-quiet" ellipticals of Phillips et al.
(1986) to the more powerful radio galaxies of McCarthy (1988), nicely covering
FRIs and FRIIs which are located within the transition region in radio power
(Zirbel \& Baum 1995).
For an FRI radio galaxy, Zirbel \& Baum (1995) defined it solely on the basis of
radio morphology as a radio galaxy in which 
the radio emission peaks near the center of the galaxy and the twin jets
fade with distance from the center producing the diffuse, plumelike and
edge-dimmed radio lobes.
Complete samples of FRI radio galaxies, such as 3CR (Laing et al 1983),
B2 (Ulrich 1989), Wall-Peacock (Wall \& Peacock 1985),
and other samples are all included in the Zirbel-Baum sample.
It is therefore thought that although the Zirbel-Baum FRI sample itself is not complete,
it can represent the FRI population (e.g. Kollgaard et al. 1996; Urry \& Padovani 1995).
Some sources have been reclassified. For
example, 3C 28, 3C 293, 3C 305, 3C 310, 3C 346, 3C 433 and NGC 6109 which were
identified to be FRIs by Laing et al. (1983) have been reclassified as FRIIs,
3C 40 which was identified to be an FRII by Morganti et al. (1993) has 
been reclassified as an FRI, and
Caganoff's sources (Caganoff 1988) have been all reclassified from the
original radio maps (Zirbel \& Baum 1995).
Although it is true that some of FRII radio galaxies may also be parent population
of BL Lac objects which are probably identical to LBLs,
we exclude possible FRIIs which have been identified as FRIs
initially but have been reclassified as FRIIs later, since our purpose is
to investigate whether there exist HBL-like and LBL-like subclasses of FRI
radio galaxy.
For some sources, X-ray data are not available at present. These sources are
therefore excluded from our sample, yet our sample
comprises over 70\% FRI radio galaxies in the Zirbel-Baum sample.

In Table 1, the columns are: (1) IAU name; (2) other name; (3) redshift; 
(4) flux density of the core at 5GHz in unit of mJy; (5) reference 
for the radio flux density; (6) flux density of the central unresolved
component at 1keV in unit of nJy; (7) reference for X-ray flux;
(8) effective spectral indices between 5GHz and 1keV corrected for beaming.
For some objects, X-ray flux densities
are calculated from broad band flux with the photon index (or spectral
index) given in the corresponding reference.
Effective spectral index $\alpha_{rx}$ between 5GHz and 1keV is calculated as
(Lamer et al. 1994)
$$
\alpha_{rx} = \frac{\log_{10}(S_{5GHz}/S_{1keV})}{7.68},
$$
and between 5GHz and 2keV as (Perlman et al. 1996)
$$
\alpha_{rx} = \frac{\log_{10}(S_{5GHz}/S_{2keV})}{7.985},
$$
where $S_{5GHz}$, $S_{1keV}$ and $S_{2keV}$ are flux densities at 5GHz, 1keV
and 2keV in the rest frame of the source, respectively. The flux densities
in Table 1 have been all $K$-corrected before computing $\alpha_{rx}$.
Since the radio cores of FRIs have a flat spectrum, we take $\alpha_{r}$=0.0
for the $K$-correction when there is no $\alpha_{r}$ in the literature,
and $\alpha_{x}$=0.8, the typical value for FRI class.

The value of $\alpha_{rx}$ in Table 1 has been corrected for beaming as well.
According to the unified scheme, an FRI radio galaxy with $\alpha_{rx}^{F}$ is
viewed at a large angle to the line of sight. When viewed at a small angle,
it will appear as a BL Lac object with $\alpha_{rx}^{B}$.
The transformation law for the change of $\alpha_{rx}$ due to relativistic
beaming is given by Chiaberge et al. (2000) as
$$
\alpha_{rx}^{B}=\alpha_{rx}^{F} + (\alpha_{r}-\alpha_{x})
\frac{\log_{10}(\delta_{B}/\delta_{F})}{7.68},
$$
where $\delta_{B}$ and
$\delta_{F}$ are Doppler factors for BL Lac objects and FRI radio galaxies,
respectively. The Doppler factor is defined as
$\delta = [\Gamma(1-\beta\cos\theta)]^{-1}$, where $\beta$ is the speed
of emitting plasma in unit of the speed of light, $\Gamma$ is Lorentz factor
$\Gamma=(1-\beta^2)^{-1/2}$ and $\theta$ is the angle to the line of sight.
We take typical value for viewing angle and Lorentz factor.
The typical viewing angle of FRI radio galaxies is $\theta=60\arcdeg$,
and for BL Lac objects, the typical viewing angle is $\theta=20\arcdeg $
(e.g. Padovani 1999).
The typical Lorentz factor for relativistic jets is $\Gamma=5.0$ 
(Padovani 1999; Param et al. 1996; Marscher 1993).

\section{DISCUSSION AND CONCLUSIONS}

As mentioned in the Introduction section, a practical way to parameterize
the different SEDs is to use the radio-to-X-ray spectral index, $\alpha_{rx}$,
and according to whether $\alpha_{rx}$ is greater than or less than 0.75
(Padovani \& Giommi 1995)
BL Lac objects are divided into two subclasses, LBL and HBL. 
Figure 1a displays the distribution of $\alpha_{rx}$ for our sample of 51 
FRI radio galaxies.
For comparison, Figure 1b illustrates the distribution of $\alpha_{rx}$
for BL Lac objects in the samples of EMSS (the most X-ray-dominated BL Lac
objects), $ROSAT$-Green Bank (RGB, intermediate) and 1Jy (the most
radio-dominated BL Lac objects, see the fourth paragraph of this section
for detailed properties of BL Lac samples).
For RGB objects, $\alpha_{rx}$ is calculated with the data in 
Laurent-Muehleisen et al. (1999), and for 1Jy and EMSS objects,
$\alpha_{rx}$ is taken from Sambruna et al. (1996).
We adopt the data in the RGB sample for overlapping sources between the RGB and EMSS
or 1Jy sample.
Important features are seen in Figure 1a and Table 1:
a) The range of $\alpha_{rx}$ of FRI radio galaxies is 0.45 to 0.92,
which is almost the same as that of BL Lac objects (0.45 to 1.01,
see Figure 1b);
b) There is a peak at $\alpha_{rx}\approx$0.70 and the average of
$\alpha_{rx}$ for the whole sample is 0.70. (The
average of $\alpha_{rx}$ of BL Lac objects in Fig. 1b is 0.69.);
c) 38 out of total 51 FRIs have $\alpha_{rx}<0.75$.
These suggest that HBL-like FRI radio galaxies are not
rare exceptions as in the case for flat spectral radio quasars
(FSRQs; Padovani, Giommi, \& Fiore 1997),
but statistically constitute a subclass which populates about half of the sample.
These results shows that there exist two subclasses of FRI radio galaxies:
one is HBL-like and the other is LBL-like.

Investigations of multifrequency spectral energy distribution from the radio
through the X-ray bands for BL Lac objects and FSRQs show that the different
SEDs between HBL and LBL cannot be explained in terms of different viewing
angles alone (Sambruna et al. 1996; Fossati et al. 1998).
There must be some intrinsic differences that cause different
SEDs between LBL and HBL. As their parent population, FRI radio galaxies
should exhibit similar differences,
but should not be what was once believed that every FRI could appears as an
HBL or an LBL if viewed at different angles to the line of sight.
Therefore, the existence of the HBL-like and LBL-like subclasses of  FRI 
radio galaxies supports the unified scheme of BL Lac objects and FRI radio galaxies.

In fact, as early as in 1992, based on VLA observation for complete samples
of LBL and HBL, Kollgaard et al. (1992) and Laurent-Muehleisen et al. (1993)
found that only those FRI radio galaxies with
intrinsically stronger cores would appear as an LBL if viewed at a small angle
to the jet direction, while HBL may be found among the weaker cored FRI radio
galaxies. This conflicted with the prevailing point of view at that time that
XBLs (mainly HBLs) were intrinsically the same 
as RBLs (mainly LBLs), but now has actually become another evidence
of the unified scheme, and is consistent with our results. 
In addition, the result of Kollgaard et al. also suggests that
care should be taken when using flux-limited complete samples to test the unified scheme,
since there may be few HBL-like FRIs in the samples of brighter radio source.

In Fig. 1a, there is only one peak at $\alpha_{rx}\approx 0.72$ in the
distribution of $\alpha_{rx}$. It had been known for long that the
distribution of BL Lac objects is bimodal which peaks at
$\alpha_{rx}\approx0.5$ and 0.9 (e.g. Stocke et al. 1985, 1991; Padovani \&
Giommi 1995, 1996; Sambruna et al. 1996; Fossati et al. 1998).
However, a large fraction of newly discovered BL Lac objects have been found to fill in the gap
between the two peaks, which are called intermediate BL Lac objects (IBLs,
Perlman et al. 1998; Caccianiga et al. 1999; Laurent-Muehleisen 1997; Laurent-Muehleisen et al. 1999).
Probably, as Laurent-Muehleisen et al. (1999) pointed out,
the bimodal distribution of BL Lac objects found in earlier studies
was caused by observational selection effects, and the true distribution of
X-ray-to-radio flux ratio $S_{x}/S_{r}$ (equals the distribution of $\alpha_{rx}$) is unimodal
(Laurent-Muehleisen et al. 1999).
A Kolmogorov-Smirnov test shows that the distributions of $\alpha_{rx}$ for
FRI radio galaxies and BL Lac objects in Figs. 1a and 1b are indistinguishable
at the level of 90 percent.
However, further quantitative test is necessary in the future,
because it seems that none of the flux-limited complete samples of
BL Lac objects available at present can represent the whole class of BL Lac
objects.

Recently, Chiaberge et al. (2000) tested the unification of BL Lac objects
and FRI radio galaxies by comparing the core emission of radio galaxies with
those of BL Lac objects, taking advantage
of the $HST$ optical observation for the core in 3CR FRI radio galaxies.
From the comparison of the optical and radio emission of FRIs and BL Lac
objects,
they inferred Lorentz factors of $\Gamma \sim 5$, which are typical for
a relativistic jet (Padovani 1999; Param et al. 1996; Marscher 1993),
supporting the unified scheme.
Their study also shows that in the $L_{o}-L_{r}$ (optical and radio luminosities
of the core) plane, HBL and LBL move on different debeaming trails, 
and that debeaming trail of LBL crosses the 3CR FRI region (both the simple emission
component model and two-velocity jet model).
By taking the typical value for the Lorentz factor or
assuming a two-velocity structures for the jet, the location of debeamed
BL Lac objects in the 1Jy sample in the $L_{o}-L_{r}$ plane coincides with
that of FRIs in the 3CR sample,
indicating that FRIs in the 3CR sample are LBL-like or IBL-like.
This is also consistent with our result. 

It is notable that the percentage of LBL-Like FRIs in our sample is less than
50\% after beaming correction. This may indicate that some FRIIs or at least
the intermediate sources (FRI/II) should be included in the parent population
of BL Lac objects, which was also suggested by VLA observations (Kollgaard
et al. 1992) and optical spectroscopic observations 
(Owen et al. 1996). Another possible explanation is that this may be caused
by the limited angular resolution of X-ray observations. For example,
the angular resolution of $ROSAT$ with the HRI is only about 5$\arcsec$.
The X-ray flux of the central unresolved component of an FRI radio galaxy is
probably not contributed by the VLA radio core alone. In the near future, 
Chandra X-ray Observatory (CXO) will observe the central region of FRI radio
galaxies with much higher spatial resolution and sensitivity,
obtaining a much better and larger X-ray-observed sample to test
the unified scheme more reliably.
In addition, with the CXO's high sensitive and high spectral resolution observation,
the existence of HBL-like and LBL-like FRI radio galaxies can be tested directly by
comparing the X-ray spectra. Since the HBL-like FRIs peak the synchrotron
emission at UV/X-ray energy range, the X-rays are an extension of synchrotron
emission with a steeper spectrum, and the shape of the optical-to-X-ray
continuum is convex ($\alpha_{x}>\alpha_{ox}$); while for most LBL-like FRIs,
the X-rays are dominated by inverse-Compton emission with a flatter spectrum, 
and the shape of the optical-to-X-ray continuum is concave ($\alpha_{x}<\alpha_{ox}$).
Therefore, HBL-like FRIs can be distinguished from LBL-like FRIs in the
optical to soft X-ray energy range.

In summary, the range of $\alpha_{rx}$ for FRI radio galaxies is almost the
same as that for BL Lac objects, the distribution of $\alpha_{rx}$ probably
peaks at the same position as BL Lac objects, and the distribution of
$\alpha_{rx}$ for FRIs is similar to that for BL Lac objects. These suggest
that there exist two subclasses of FRI radio galaxies: one is HBL-like, and
the other is LBL-like, corresponding to HBLs and LBLs, respectively. This result
supports the unified scheme of BL Lac objects and FRI radio galaxies.

\acknowledgments

The authors wish to thank the anonymous referee for his/her comments that
improved this paper significantly.
This work was financially supported by the BK21 Project
of the Korean government.

\input{epsf}
\begin{figure}[htp]
\vskip -15.5cm
\begin{center}
\hskip 0.1cm
\epsfysize=3in\epsfxsize=4.17in
\epsffile{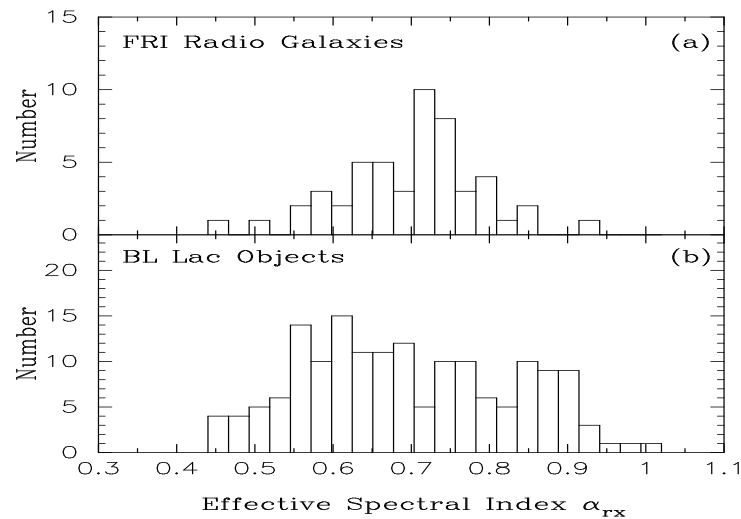}
\end{center}
\vskip -0.5cm
\caption { Distributions of the effective spectral index, $\alpha_{rx}$.
(a): for our sample of FRI radio galaxies,
(b): for BL Lac objects in the RGB, 1Jy and EMSS samples.}
\end{figure}


\end{document}